\documentclass[%
 preprint, 
 amsmath,amssymb,
 aps, physrev,
showkeys
]{revtex4-2}

\usepackage{graphicx}
\usepackage{dcolumn}
\usepackage{bm}
\usepackage{hyperref}

\usepackage{graphicx, subcaption}
\usepackage{lipsum}
\usepackage{tensor}
\usepackage{amsmath}
\DeclareMathOperator{\Tr}{Tr}

\usepackage{soul}
\usepackage{xcolor}

\definecolor{ifblue}{RGB}{0, 51, 102}
\definecolor{ifred}{HTML}{61172b}
\definecolor{ifgreen}{RGB}{0, 180, 0}

\newcommand{\p}[1]{
\left({#1}\right)}

\begin{document}


\title{\textbf{ Hawking-Page transition in anti-de Sitter massive gravity  with non-compact spatial boundary} 
}%

\author{Nelson R. F. Braga}
\email{braga@if.ufrj.br}
\affiliation{Instituto de Física, Universidade Federal do Rio de Janeiro, Caixa Postal 68528, RJ 21941-972, Brazil.}

\author{William S. Cunha}
\email{wscunha@pos.if.ufrj.br}
\affiliation{Instituto de Física, Universidade Federal do Rio de Janeiro, Caixa Postal 68528, RJ 21941-972, Brazil.}


\date{\today}

\begin{abstract}
In gauge/gravity duality, the Hawking Page (HP) process plays the important role of representing a confined/deconfined phase transition. For spaces that are asymptotically anti-de Sitter (AdS) with planar spatial boundary there is no HP transition, unless some infrared energy scale is introduced. 
Massive gravity theories have been applied in the context of holography 
in order to represent systems with  momentum dissipation. In this work we study the HP transition for massive gravity in an asymptotically AdS space with $\mathbb{R}^n\times\,  $S$^1 $  boundary. It is shown that the graviton mass term produces an HP transition at a finite critical temperature.

\end{abstract}

\maketitle
\newpage

\section{\label{sec:level1} Introduction}

 The Hawking-Page (HP) transition  was proposed in  \cite{Hawking:1982dh},  using a semiclassical approach to the thermal behavior of Euclidean 4-d spaces with negative cosmological  constant and spherical symmetry. There are two solutions for the  Einstein's equations under these conditions: the thermal anti-de Sitter (\(AdS\))   and the black hole anti-de Sitter  (\textit{AdS-BH}) space-times. At low temperatures, the black hole solution is unstable, so the system is in the \(AdS\) phase. For temperatures higher than some critical value, the situation  change and the system is dominated by the \textit{AdS-BH} solution. 

 In the context of gauge/string duality, it was proposed by Witten  \cite{Witten:1998zw} that the HP transition in $AdS_5$ space  corresponds to a confinement/deconfinement transition in the 4-d gauge theory side of the $AdS/CFT$ correspondence. The transition analyzed in \cite{Witten:1998zw} occurs only in the case when the $AdS_5$ space has a compact boundary. 
 For the $AdS_5$ space with planar boundary, more appropriate to represent our physical world, there is no  HP transition and the black hole is the stable configuration at any temperature. Unless some modification is made in order to introduce some dimensionfull parameter in the background, in general associated with an infrared cutoff, as it is done in \cite{Herzog:2006ra, BallonBayona:2007vp,Colangelo:2009ra,Colangelo:2011sr, 
 PhysRevD.108.114020,Braga:2024nnj,Braga:2025eiz}.

Massive gravity theories, meaning theories where the graviton is massive,  have been studied for quite a long time (see for example \cite{deRham:2014zqa,Hassan:2011vm}). In the gauge/gravity dualities they are interpreted as representing systems with broken translational symmetry and, consequently, momentum dissipation \cite{Vegh:2013sk}. The HP transition in the case of $AdS$ space with spherical boundary with the addition of a graviton mass term was studied in \cite{Adams:2014vza}. It was found that the presence of the mass term in the gravitational action lowers the critical temperature, with respect to the non massive case. 
    
 In this work, we consider $AdS$ space with planar boundary. As mentioned before, in this case there is no HP transition if one considers just the standard Einstein action for this space. We will introduce a graviton mass term in the action and then analyze the thermodynamics of the resulting system. A phase transition at a finite temperature, related to the mass scale of the graviton, will show up.

The paper is organized as follows. In Sec. \ref{sec: 2}, we review some illustrating examples of Hawking-Page transitions. In Sec \ref{sec: 3}, we  present some aspects of massive gravity that are relevant to this work. In Sec. \ref{sec: 4} we discuss the stability of the black hole solution in the massive AdS space and then examine the HP transition between the spaces with and without a black hole, focusing on the system's thermodynamics. Finally in Sec.\ref{sec: conclusion} our conclusions and final remarks are presented. The detailed computations for the metrics used in this work are provided in Appendix \ref{sec:Ap einstein equations}.

\section{\label{sec: 2} Hawking-Page transition in $AdS$ space}

    We begin our discussion by revisiting the concept of Hawking Page transition, originally proposed for Euclidean anti-de Sitter space with periodic time coordinate in \cite{Hawking:1982dh}. The main idea is that in a quantum theory of gravity the more relevant contributions to the 
    path integral, that sums over all possible metric configurations that are asymptotically $AdS$,  comes from the corresponding classical solutions of Einstein equations. There are two solutions, one with a black hole and the other without. The dominant (or stable) is the one with the smallest action integral.  This condition mirrors the thermodynamic stability criterion of minimizing the free energy.

    \subsection{The $ AdS_{_D}$ space  with $ S^{^{D-2}}\times S^1$  boundary}

    Let us extend the case of spacetime dimension \(D=4\) considered in \cite{Hawking:1982dh} to arbitrary dimensions. The Einstein Hilbert action has the form
        \begin{equation}\label{eq: ads action}
            \mathcal{I} = -\frac{1}{2\kappa^2}\int d^{^{D}}x\sqrt{ \vert g \vert }\p{\mathcal{R}-\Lambda}
        \end{equation}
where \(\kappa^2 =8\pi G_{_D}\) with \(G_{_D}\)  the \(D\)-dimensional Newton's constant with dimension \(E^{-(D-2)}\), \(\mathcal{R}\) and  \(\Lambda\) denote respectively the Ricci scalar and cosmological constant. For \(AdS\) spaces with radius \(L\), they depend on the spacetime dimension \(D\) as follows
        \begin{equation}\label{eq: ricci and cosmological constant}
            \mathcal{R} = - \frac{D(D-1)}{L^2} \quad \text{and} \quad \Lambda=-\frac{(D-1)(D-2)}{L^2}\,.
        \end{equation}
        
    The black hole metric with spherical symmetry that emerges as a solution of the action \eqref{eq: ads action} with imaginary time is
        \begin{equation}\label{eq: metric of ads spheric}
            ds^2 = g_{pq}dx^pdx^q =  V(r)dt^2 + V(r)^{-1}dr^2 +r^2 d\Omega_n^2\,,
        \end{equation}
    here \(n = D - 2\) is the number of spatial dimensions in the boundary, and \(d\Omega_n^2\) represents the metric of the \(n\)-dimensional sphere. The function \(V(r)\) takes the form
        \begin{equation}
            V(r) = 1-\frac{M}{r^{n-1}}+\frac{r^2}{L^2}
        \end{equation}
    with \(M\) being proportional to the black hole mass. The horizon  position $ r_h$ is determined by $V (r_h) = 0 $ and the Hawking temperature is obtained requiring that there is no conical singularity in the vicinity of the horizon. The result is
        \begin{equation}
            T = \frac{(n-1)L^2+(n+1)r_h^2}{4\pi L^2 r_h}\,.
        \end{equation}
        
      For the thermal $AdS $ space the metric in the same form of \eqref{eq: metric of ads spheric}, but with \(M=0\), i.e. 
        \begin{equation}
            V_0(r)  = 1+\frac{r^2}{L^2}.
        \end{equation}

        From Eqs.  \eqref{eq: ads action} and  \eqref{eq: ricci and cosmological constant} one obtains the actions for the black hole \(AdS\)
        \begin{equation}
            \mathcal{I}^{(BH)} =\frac{\Omega_n}{\kappa^2}\frac{\beta}{L^2}\left(r_b^{n+1}-r_h^{n+1}\right)\,,
        \end{equation}
        and for the thermal \(AdS\)
        \begin{equation}
            \mathcal{I}^{(AdS)} =\frac{\Omega_n}{\kappa^2}\frac{\beta_{0}}{L^2}r_b^{n+1}\,.
        \end{equation}
        Here, \(r_b\) denotes a regulator for the boundary position, \(\Omega_n\) is the area of the \(n\) sphere, while \(\beta\) and \(\beta_{0}\) originate from the temporal coordinate integration. These quantities can be related by requiring that the two geometries have the same temporal length  at the boundary, yielding
        \begin{align}
            \beta_{0} &= \left(\frac{V(r_b)}{V^{'}(r_b)}\right)^{1/2}\beta\nonumber\\
            &\approx \p{1- \frac{L^2 M}{2 r_b^{n+1}}}\beta\,,
        \end{align}
        where the approximation  holds for large \(r_b\).

        To remove the divergences in the limit \(r_b\to \infty\), one can define the renormalized action difference as
        \begin{align}
            \Delta \mathcal{I} &\equiv \mathcal{I}^{(BH)} - \mathcal{I}^{(AdS)}\nonumber\\
            &=\frac{\Omega_n}{2\kappa^2}\beta \p{r_h^{n-1}-\frac{r_h^{n+1}}{L^2}},
        \end{align}
        where the condition $ V(r_h) = 0$ was used.

        As explained in the beginning of this section, the stable space configuration corresponds to the lowest action. Therefore, for \(\Delta\mathcal{I}>0\) the thermal \(AdS\) space is stable, for \(\Delta\mathcal{I}<0\) the black hole solution becomes the stable one. The phase transition occurs at \(\Delta\mathcal {I} =0 \), where both configurations have equal action values. This condition determines the critical temperature for the Hawking-Page transition as
        \begin{equation}
            T_c = \frac{n}{2\pi L}.
        \end{equation}

        
    \subsection{ The \(AdS_{_D}\) space  with \(R^{^{D-2}} \times S^1 \) boundary}

        It was proposed in \cite{Witten:1998zw}, in the context of gauge  gravity duality,  that the Hawking-Page transition in the gravitation side corresponds to the confinement/deconfinement phase transition in the gauge theory side. 
    
        In order to represent a gauge theory living in a non-compact space, one considers a flat \(R^{^{D-2}}\) spatial boundary geometry. The metric of the black hole in Poincaré coordinates becomes
        \begin{equation}
            ds^2 = \frac{L^2}{z^2}\p{ f(z) dt^2 + f(z)^{-1}dz^2 + d\vec x_n^2},
        \end{equation}
        where the horizon function \(f(z)\) is given by
        \begin{equation}
            f(z) = 1-\frac{z^{n+1}}{z_h^{n+1}}\,.
        \end{equation}
        The condition of absence of conical singularity leads to the Hawking temperature
        \begin{equation}\label{eq: temperature as zh}
            T = \frac{(n+1)}{4\pi}\frac{1}{z_h}.
        \end{equation}                

        The metric of the thermal \(AdS\) is obtained using \(f(z)=1\). Following the same procedure as in the case of compact spatial boundary, one obtains the action difference
        \begin{equation}\label{eq: action to planar ads}
            \Delta\mathcal{I}=-\frac{L^n}{2\kappa^2}\beta V_n \frac{1}{z_h^{n+1}}
        \end{equation}
        where \(V_n\) is the \(n\)-dimensional spatial volume. 

        Equation \eqref{eq: action to planar ads} is zero only when \(z_h\to\infty\) and the temperature goes to zero by the equation \eqref{eq: temperature as zh}. For any non-zero temperature \(\Delta\mathcal{I} <0 \). So, the black hole geometry is always the stable one. In other words, there is no HP transition.

        The zero critical temperature is a consequence of the absense of any dimensionful parameter in the  gauge string duality in this case.   Important   physical systems governed by gauge theories, like the quark gluon plasma (QGP), typically exhibit a finite critical temperature. In order to model holographically such systems, various approaches have been proposed that modify the \(AdS\) geometry in order to to break conformal invariance in the gauge theory side, thereby generating non-zero critical temperatures. Some examples can be found in \cite{Herzog:2006ra, BallonBayona:2007vp,Colangelo:2009ra,Colangelo:2011sr,  PhysRevD.108.114020,Braga:2024nnj,Braga:2025eiz}. In the following section we will study a different situation where an 
        energy parameter is naturally present: the case when the gravitons are massive.

\section{\label{sec: 3}   AdS space with Massive Gravity}

    The formulation of massive gravity that we use in this work was proposed in \cite{deRham:2010ik, deRham:2010kj, Hassan:2011vm} and used in \cite{Adams:2014vza} to analyze the HP transition for the case of spherical spatial boundary. One introduces a graviton mass term in the action of the form
    \begin{equation}\label{eq: gravtion action}
        {\cal I}_{\text{grav.}}=-\frac{1}{2\kappa^2}\int d^{n+2}x\sqrt{\vert g \vert } \; m^2\sum^{n+2}_{i=1}c_i{\cal U}_i
    \end{equation}        
    where \(m\) is the graviton mass, \(c_i\) are dimensionless constants and \({\cal U}_i\) are the symmetric polynomial eigenvalues of the matrix \({\cal K }\). We consider only the first two terms of the sum:
    \begin{align}
        {\cal U}_1 &= \Tr{\cal K},\label{U1}\\
        {\cal U}_2 &= \p{\Tr{\cal K}}^2-\Tr{\cal K}^2. \label{U2}
    \end{align}
    
    The  \((n+2)\times(n+2)\) matrix \({\cal K}\) is defined in terms of the dynamical metric  and of a reference metric \(f_{\mu\nu}\) as follows
    \begin{equation}\label{eq: matrix k definition}        \mathcal{K}\indices{^\mu_\rho}\mathcal{K}\indices{^\rho_\nu}=g^{\mu\sigma}f_{\sigma\nu}
    \end{equation}

    Following \cite{Vegh:2013sk, Blake:2013bqa}, we consider  the case of a flat reference metric
    \begin{equation}\label{eq: refence metric}
        f_{\mu\nu} = \sum_j\delta^j_\mu\delta_\nu^j\,,
    \end{equation}
    where the sum over \(j\) runs over the spatial boundary coordinates.
    
    This choice breaks the diffeomorphism invariance in the spatial coordinates of the boundary, leading to non-conservation of spatial momentum in the boundary theory. For more details about massive gravity theory see the reviews \cite{deRham:2014zqa, Hinterbichler:2011tt}.

    Considering that only \(c_1\) and \(c_2\) are non-zero in the graviton action \eqref{eq: gravtion action}, one can write an effective action as
    \begin{equation}\label{eq: ads graviton action}
        {\cal I}=-\frac{1}{2\kappa^2}\int d^{n+2}x\sqrt{ \vert g \vert}\p{\mathcal{R}-\Lambda}-\frac{1}{2\kappa^2}\int d^{n+2}x\sqrt{\vert g \vert}\p{\lambda_1\Tr{\cal K} +\lambda_2 \left[ \p{\Tr{\cal K}}^2 -\Tr{\cal K}^2\right ]},
    \end{equation}
    where we have defined \(\lambda_1\equiv c_1 m^2\) and \(\lambda_2\equiv c_2m^2\) for simplicity, both parameters having dimensions of energy squared.
    
    The black hole metric solution of the action \eqref{eq: ads graviton action} that is asymptotically \(AdS\) and has planar symmetry in the spatial coordinates has the form
    \begin{equation}\label{eq: MG bh metric}
        ds^2 = \frac{L^2}{z^2}\left( f(z)dt^2 +f(z)^{-1}dz^2+ d\vec x_n^2\right),
    \end{equation}
    with a modified horizon function given by
    \begin{equation}\label{eq: MG horizon function}
        f(z) = 1 + \frac{\lambda_1 L}{n}z+\lambda_2 z^2  - \frac{z^{n+1}}{z_h^{n+1}}\p{1+ \frac{\lambda_1 L}{n}z_h+\lambda_2 z_h^2}.
    \end{equation}        
    The derivation of this solution is presented in Appendix \ref{sec:Ap einstein equations}. The Hawking temperature is obtained  by requiring regularity in imaginary time, giving
    \begin{equation}\label{eq: MG temperature}
        T = \frac{1}{4\pi z_h}\p{\p{n+1}+\lambda_1 L z_h + \p{n-1}\lambda_2 z_h^2},
    \end{equation}
    remembering that \(z_h\) denotes the horizon position.  The solution without a black hole corresponds to the limit \(z_h\to\infty\), which is equivalent to setting the black hole mass to zero, this leads to 
    \begin{equation}\label{eq: MG function}
        f_0(z) = 1 + \frac{\lambda_1 L}{n}z+\lambda_2 z^2 .
    \end{equation}

    \section{\label{sec: 4}HP transition in Massive AdS space} 

    \subsection{Regularized action}
   
        Let us first calculate the gravitational term of action \eqref{eq: ads graviton action}. Using the same cosmological constant from Eq. \eqref{eq: ricci and cosmological constant}, expressed in terms of \(n\), one obtains:
        \begin{equation}\label{eq: first term of MG action}
            {\cal I}_1 = -\frac{1}{2\kappa^2}\int d^{n+2}x\sqrt{ \vert g \vert }\p{\mathcal{R}+\frac{n(n+1)}{L^2}}\,.
        \end{equation}
        
        The additional terms in the action \eqref{eq: ads graviton action} modify the Ricci scalar, which becomes 
        \begin{equation}\label{eq: ricci correct}
            {\cal R} = -\frac{(n+1)(n+2)}{L^2}-\frac{(n+1)\lambda_1 L z}{L^2} -\frac{n(n-1)\lambda_2 z^2}{L^2}\,,
        \end{equation}
        using this result in equation \eqref{eq: first term of MG action} yelds
        \begin{equation}
            {\cal I}_1 = \frac{L^n}{2\kappa^2}\beta V_n\int^{z_h}_\varepsilon dz \p{\frac{2(n+1)}{z^{n+2}}+\frac{(n+1)\lambda_1 L}{z^{n+1}}+\frac{n(n-1)\lambda_2}{z^n}}\,, 
        \end{equation}
        where \(V_n\) and \(\beta\) originate from the integration over the boundary coordinates and \(\varepsilon\) is a regularization parameter. Integrating in the holographic coordinate one finds:
        \begin{align}
            {\cal I}_1 = \frac{L^n}{\kappa^2}\beta V_n\left(- \frac{1}{z_h^{n+1}}-\frac{(n+1)}{2n}\frac{\lambda_1 L}{z_h^n} -\frac{n}{2}\frac{\lambda_2}{z_h^{n-1}}\right. 
            + \left. \frac{1}{\varepsilon^{n+1}}+\frac{(n+1)}{2n}\frac{\lambda_1 L}{\varepsilon^n} +\frac{n}{2}\frac{\lambda_2}{\varepsilon^{n-1}}\right).
        \end{align}

        Now let us consider  the term linear in the trace of \({\cal K}\) in the action \eqref{eq: ads graviton action} 
        \begin{equation}\label{eq: second term of MG action}
            {\cal I}_2 = -\frac{1}{2\kappa^2}\int d^{n+2}x\sqrt{ \vert g \vert } \, \lambda_1 \Tr{\cal K}\,.
        \end{equation}        
        Using the definition \eqref{eq: matrix k definition} and the metric \eqref{eq: MG bh metric}, the trace reduces to
        \begin{equation}\label{eq: trace of k}
            \Tr{\cal K} = n \frac{z}{L}\,.
        \end{equation}
        Consequently, the action \eqref{eq: second term of MG action} becomes 
        \begin{align}
            {\cal I}_2 &= -\frac{1}{2\kappa^2}\beta V_n \int^{z_h}_\varepsilon dz\frac{L^{n+2}}{z^{n+2}}\lambda_1 \,n \frac{z}{L} \nonumber\\
            &=\frac{1}{\kappa^2}\beta V_n \p{\frac{\lambda_1 L}{2z_h^n}- \frac{\lambda_1 L}{2\varepsilon^n}}.
        \end{align}

The last term in the massive gravity action consists of the quadratic combination 
        \begin{equation}\label{eq: third term of MG action}
            {\cal I}_3 = -\frac{1}{2\kappa^2}\int d^{n+2}x\sqrt{ \vert g \vert }\, \lambda_2\left[ \p{\Tr{\cal K}}^2 -\Tr{\cal K}^2\right ].
        \end{equation}
        For the diagonal matrix \({\cal K}\), this combination simplifies to 
        \begin{equation}
            \left[ \p{\Tr{\cal K}}^2 -\Tr{\cal K}^2\right ] = n\p{n-1}\frac{z^2}{L^2}\,.
        \end{equation}
        Inserting this result into \eqref{eq: third term of MG action} yields
        \begin{align}
            {\cal I}_3 &= -\frac{1}{2\kappa^2}\beta V_n \int^{z_h}_\varepsilon\frac{L^{n+2}}{z^{n+2}}\lambda_2 \,n(n-1)\frac{z^2}{L^2}\nonumber\\
            &=\frac{L^n}{\kappa^2}\beta V_n\p{\frac{n\lambda_2}{2z_h^{n-1}}-\frac{n\lambda_2}{2\varepsilon^{n-1}}}.
        \end{align}

        Summing the three contributions yields the total action for the planar black hole solution in \(AdS\) massive gravity 
        \begin{equation}\label{eq: MG black hole action}
            {\cal I}^{(BH)} = \frac{L^n}{\kappa^2}\beta V_n\p{-\frac{1}{z_h^{n+1}}-\frac{\lambda_1 L}{2n}\frac{1}{z_h^n}+\frac{1}{\varepsilon^{n+1}}+\frac{\lambda_1 L}{2n}\frac{1}{\varepsilon^n}}.
        \end{equation}

        The action for the space without a black hole is obtained by taking the limit \(z_h\to\infty\):
        \begin{equation}
        \label{action m ads}
            {\cal I}^{(AdS)} = \frac{L^n}{\kappa^2}\beta_0 V_n\p{\frac{1}{\varepsilon^{n+1}}+\frac{\lambda_1 L}{2n}\frac{1}{\varepsilon^n}}.
        \end{equation}
        
        As in the spherically symmetric \(AdS\) case, one can relate the periods   \(\beta\) and \(\beta'\) of the time coordinates  by matching the asymptotic limits of the two geometries on the boundary
        \begin{align}
            \beta_0 &= \p{\frac{f(\varepsilon)}{f_0(\varepsilon)}}^{1/2}\beta\nonumber\\
            &\approx\p{1-\frac{\varepsilon^{n+1}}{2z_h^{n+1}}-\frac{\lambda_1 L}{2n}\frac{\varepsilon^{n+1}}{z_h^n}-\frac{\lambda_2}{2}\frac{\varepsilon^{n+1}}{z_h^{n-1}}}\beta \,, 
        \end{align}
        where the expansion about \(\varepsilon\to0\) was used. Substituting this result into the action \eqref{action m ads} yields
        \begin{equation}\label{eq: MG action without bh}
            {\cal I}^{(AdS)} = \frac{L^n}{\kappa^2}\beta V_n\p{\frac{1}{\varepsilon^{n+1}}-\frac{1}{2z_h^{n+1}}-\frac{\lambda_1 L}{2n}\frac{1}{z_h^n}+\frac{\lambda_1 L}{2n}\frac{1}{\varepsilon^n}-\frac{\lambda_2}{2z_h^{n-1}}}+{\cal O}(\varepsilon)\,.
        \end{equation}

        The regularized action is obtained through the subtraction scheme as
        \begin{align}\label{eq: regularized action}
            \Delta{\cal I}&=\lim_{\varepsilon\to0}\left( {\cal I}^{(BH)} -  {\cal I}^{(AdS)} \right) \nonumber\\
            &=-\frac{L^n}{2\kappa^2}\beta V_n\p{\frac{1}{z_h^{n+1}}-\frac{\lambda_2}{z_h^{n-1}}}.
        \end{align}

    \subsection{Stability of the black hole solution}

\begin{figure}[ht]
            \begin{subfigure}[h]{0.48\textwidth}
                \includegraphics[width=1\linewidth]{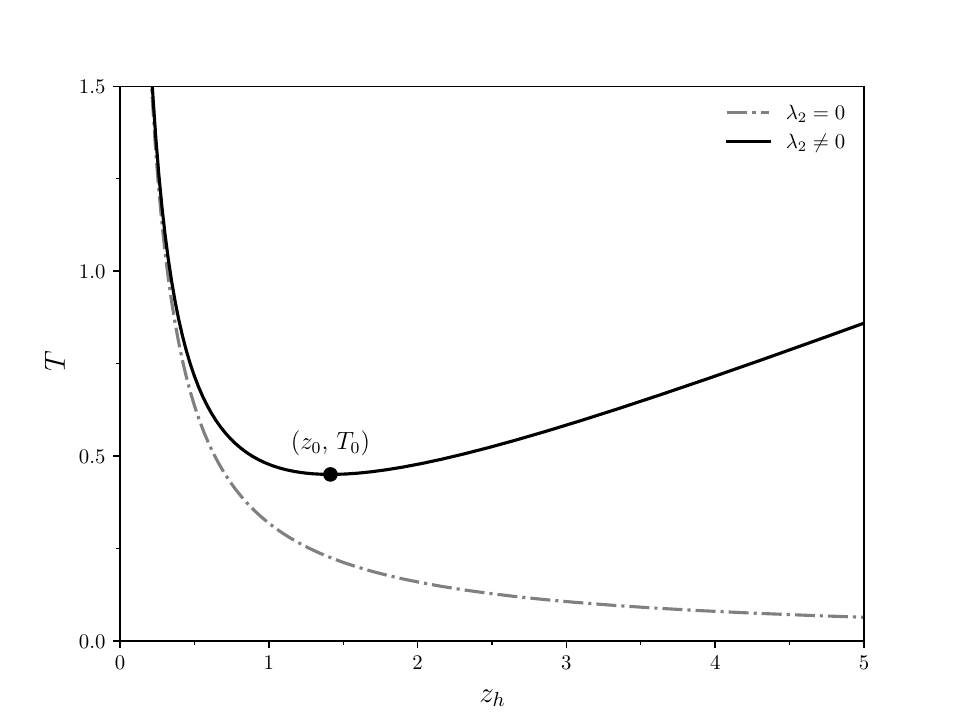}
                \caption{}
                \label{fig: MG temperature behavior}
            \end{subfigure}
            \hfill
            \begin{subfigure}[h]{0.48\textwidth}
                \includegraphics[width=1\linewidth]{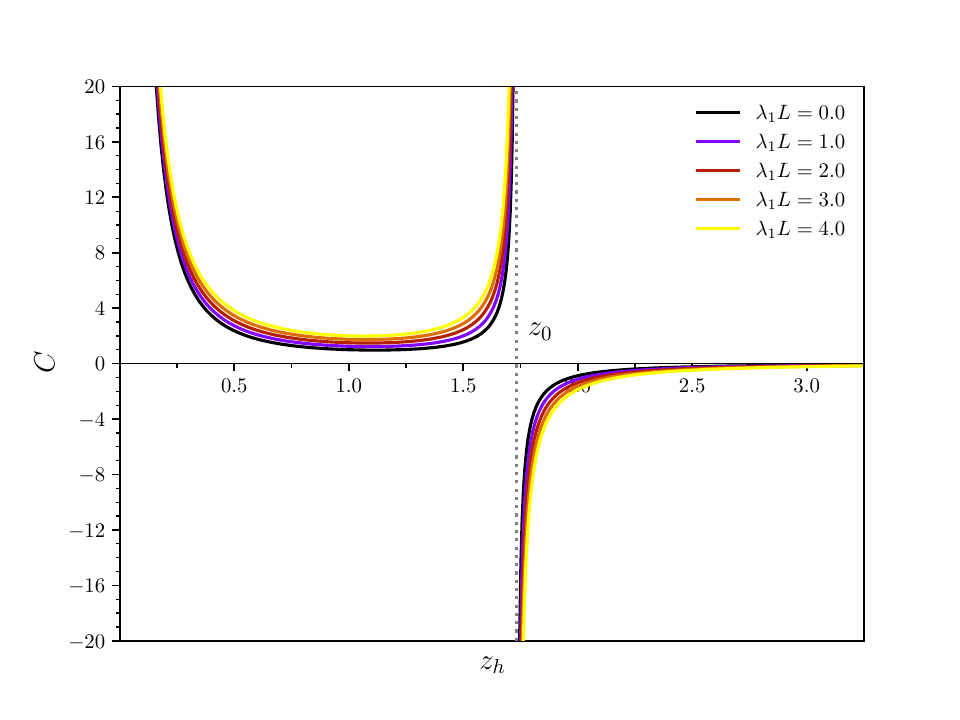}
                \caption{}
                \label{fig: Heat capacity}
            \end{subfigure}
            \caption{
                (a) Schematic plot of the temperature given by the equation \eqref{eq: MG temperature}. The dashed gray line is the case when \(\lambda_2=0\), for which temperature of the planar \(AdS\) black hole is recovered. (b) Heat capacity for various values of \(\lambda_1 L\). The dashed gray line is the \(z_0\). We set \(n=2\) and \(L^2 V_2/k^2=1\) in the plot.}
        \end{figure}

        Before we proceed to the Hawking-Page calculations, it is important to analyze the thermodynamic stability of the black hole solution. Its temperature is given by the equation \eqref{eq: MG temperature}, having two competing terms: one that decreases and another that increases with \(z_h\). The resulting variation of the temperature with \(z_h\), for $ n = 2$, is illustrated in Fig \ref{fig: MG temperature behavior}. The temperature has a minimum when the horizon position $z_h$ has the value:
        \begin{equation}\label{eq: z0 definition}
            z_0 = \p{\frac{n+1}{n-1}}^{1/2}\sqrt{\frac{1}{\lambda_2}} \,. 
        \end{equation}
        The corresponding minimum temperature that the black hole can reach is:
        \begin{equation}\label{eq: T0 temperature}
            T_0 = \frac{1}{2\pi}\p{n^2-1}^{1/2}\sqrt{\lambda_2} +\frac{\lambda_1 L}{4\pi}\,.
        \end{equation}

        Equation \eqref{eq: z0 definition} requires that \(\lambda_2\) can  take only positive real values. On the other hand, the parameter \(\lambda_1 L\) appears only in equation \eqref{eq: T0 temperature} as a  temperature shift and, in principle, can assume any real value. However, if \(\lambda_1 L\) is negative it is necessary to impose the condition
        \begin{equation}\label{eq: alpha lambda constraint}
            \lambda_1 L \ge -4\p{n^2-1}^{1/2} \sqrt{\lambda_2}\,,
        \end{equation}
        to avoid  \(T_0\) from being negative.

    The relevant quantities to analyze the black hole thermodynamic stability are the entropy and its derivative, which is related to the heat capacity. The connection with the canonical ensemble thermodynamics is established through the relation between the action and the partition function, \({\cal Z} = e^{-\Delta {\cal I}}\). Consequently
        \begin{equation}\label{eq: ln of the partition function}
            \ln{\cal Z} = -\Delta{\cal I}=\frac{L^n}{2\kappa^2}\beta V_n\p{\frac{1}{z_h^{n+1}}-\frac{\lambda_2}{z_h^{n-1}}}.
        \end{equation}

        The entropy can be expressed as 
        \begin{equation}\label{eq: entropy definition}
            S = \beta\langle E\rangle+\ln{\cal Z}
        \end{equation}
        where \(\langle E\rangle\) is the internal energy, which can be obtained by the relation
        \begin{equation}
            \langle E\rangle = - \frac{\partial}{\partial \beta}\ln{\cal Z}\,.
        \end{equation}
        Using the equation \eqref{eq: ln of the partition function}, one gets
        \begin{align}\label{eq: MG internal energy}
            \langle E\rangle &= -\frac{L^n}{2\kappa^2}V_n \left[\p{\frac{1}{z_h^{n+1}}-\frac{\lambda_2}{z_h^{n-1}}}+\beta\p{-\frac{\p{n+1}}{z_h^{n+2}}\frac{\partial z_h}{\partial\beta}+\frac{\p{n-1}\lambda_2}{z^n_h}\frac{\partial z_h}{\partial\beta}}\right]\nonumber\\
            &=\frac{L^n}{2\kappa^2}V_n \left[ \frac{n}{z_h^{n+1}}+\frac{n\lambda_2}{z_h^{n-1}}+\frac{\lambda_1 L}{z_h^{n}}\right],
        \end{align}
        where the simplification from the first to the second line is obtained using  
        \begin{equation}
            \frac{\partial z_h}{\partial \beta}=\frac{z_h}{\beta}\frac{\p{n+1}+\lambda_1 L z_h+(n-1)z_h^2}{\p{n+1}-\p{n-1}\lambda_2 z_h^2}\,,
        \end{equation}
        derived from \eqref{eq: MG temperature}. Note that this relation expresses the black hole's response to temperature changes. Another interesting observation is that besides  the geometric quantities \(L\) and \(z_h\), the parameters \(\lambda_1\) and \(\lambda_2\), and consequently the graviton mass, contribute to the system's internal energy in equation \eqref{eq: MG internal energy}.

        Substituting equations \eqref{eq: ln of the partition function} and  \eqref{eq: MG internal energy} into the entropy definition \eqref{eq: entropy definition} one finds
        \begin{align}
            S &=\frac{L^n}{2\kappa^2}\beta V_n\frac{1}{z_h^{n+1}}\left[ \p{n+1}+\lambda_1 L z_h +\p{n-1}\lambda_2 z_h^2 \right]\nonumber\\
            &=\frac{L^n}{4G_D}V_n\frac{1}{z_h^n}\,.\label{eq: MG entropy}
        \end{align}

So, one finds that the entropy is a monotonically decreasing function of the horizon position that would go to zero in the limit \(z_h\to\infty\). However, as shown in the Fig. \ref{fig: MG temperature behavior} and by equation \eqref{eq: MG temperature}, for \(z_h>z_0\) the temperature increases with $ z_h$. Therefore, in this region one would find that the  entropy would decrease with temperature. This would violate the thermodynamic laws and indicates that the black hole solutions are unstable in this region.
        
In order to illustrate the consequences of the decrease of the entropy with temperature, let us calculate the heat capacity for this system. Using the definition 
        \begin{equation}
            C = T\p{\frac{\partial S}{\partial T}}
        \end{equation}
        and equation \eqref{eq: MG entropy}, one obtains
        \begin{equation}\label{eq: MG heat capacity}
            C = \frac{nL^n}{4G_D}\frac{V_n}{z_h^n}\p{\frac{\p{n+1}+\lambda_1 L z_h+\p{n-1}\lambda_2 z_h^2}{\p{n+1}-\p{n-1}\lambda_2 z_h^2}}.
        \end{equation}

This expression is plotted in Fig. \ref{fig: Heat capacity}. From the denominator in parentheses,  one notes that the heat capacity exhibits a discontinuity at \( z_h = z_0\) where \(C\to\pm\infty\). For \(z_h>z_0\) the quadratic term in the denominator dominates, so 
\(C\) is negative in this region. This characterizes the instability of black hole solutions in this region. 
For \(z_h<z_0\), the heat capacity is always positive and the BH solutions are stable. 

Thus,  for a HP transition to occur the critical horizon position must be smaller than \(z_0\); or, in terms of the temperature, the critical temperature must be higher than \(T_0\). 
       
    \subsection{Hawking-Page Transition}
    
 As explained in the previous section, the transition occurs at \(\Delta{\cal I}=0\). For \(\Delta{\cal I}>0\) the solution without the black hole is the stable phase, while for \(\Delta{\cal I}<0\) the BH solution is stable. This condition determines, from Eq. \eqref{eq: regularized action} the critical horizon position \(z_c\) 
        \begin{gather}
            z_c =\sqrt{\frac{1}{\lambda_2}} \,. \label{eq: critical position}
        \end{gather}

\begin{figure}[ht]
            \centering
            \includegraphics[width=0.55\linewidth]{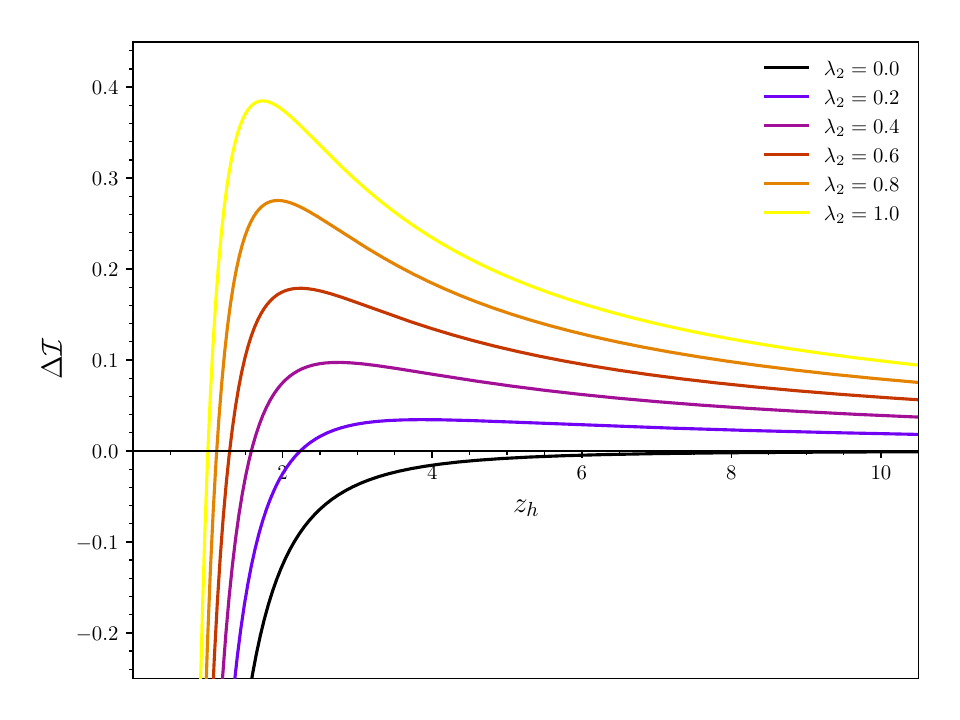}
            \caption{
                The action as a function of \(z_h\) for various values of \(\lambda_2\), the roots of the functions are the critical horizon positions \(z_c\). Was used \(n=2\) and \(L^2 V_2/k^2=1\) in the plot.}
            \label{fig: actions regularized}
        \end{figure}

This result is illustrated in Fig. \ref{fig: actions regularized}
where \(\Delta{\cal I}\) is plotted for different values of $\lambda_2$. The critical horizon positions correspond to the values of $z_h$ where the plots cross the horizontal axis. Note that $ z_c $ depends only on \(\lambda_2\), being independent of both space dimension and the parameter \(\lambda_1\). The corresponding critical Hawking-Page temperature is given by
        \begin{equation}\label{eq: critical temperature}
            T_c = \frac{n}{2\pi}\sqrt{\lambda_2} + \frac{\lambda_1 L}{4\pi}\,.
        \end{equation}

\begin{figure}[ht]
            \begin{subfigure}[h]{0.48\textwidth}
                \includegraphics[width=1\linewidth]{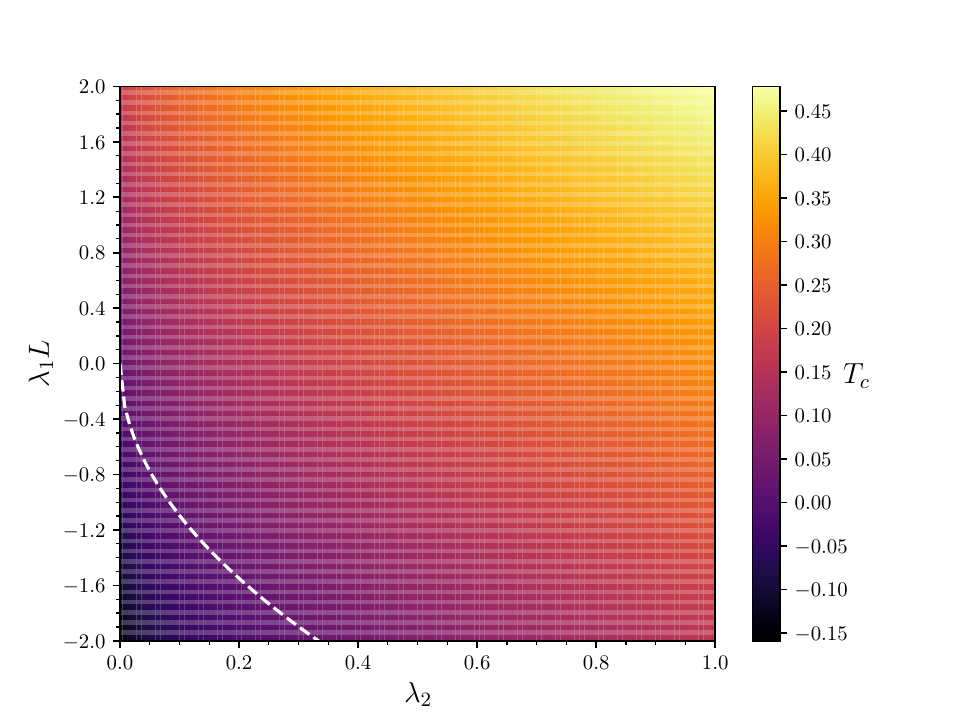}
                \caption{}
                \label{fig: Critical temperature distribution}
            \end{subfigure}
            \hfill
            \begin{subfigure}[h]{0.48\textwidth}
                \includegraphics[width=1\linewidth]{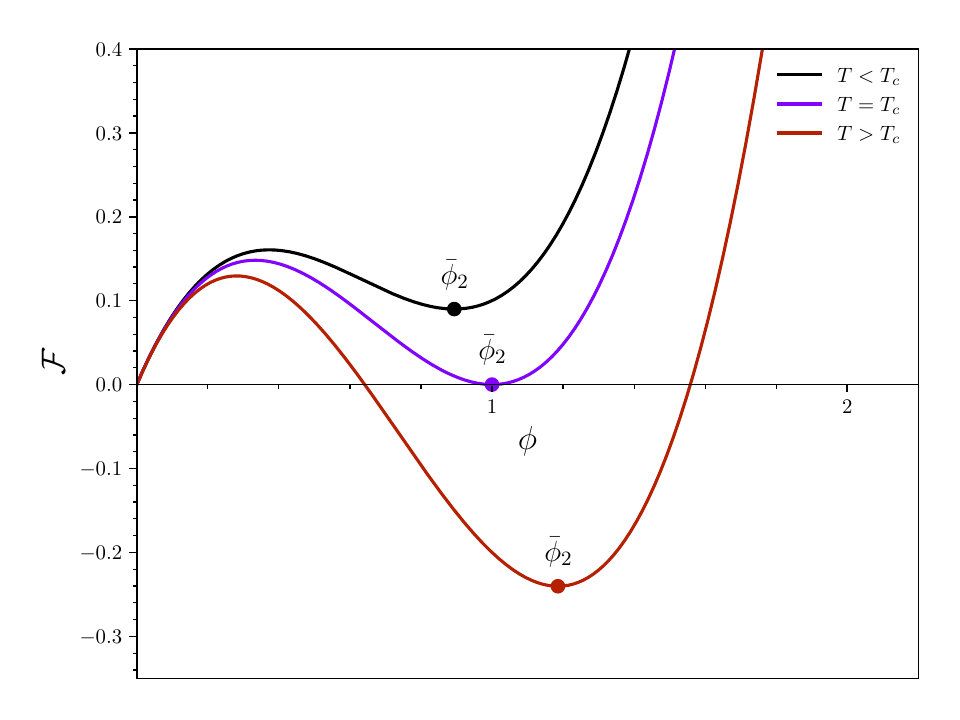}
                \caption{}
                \label{fig: landau like free energy}
            \end{subfigure}
            \caption{
                (a) Critical temperature for \(n=2\) as a function of \(\lambda_1 L\) and \(\lambda_2\). The white dashed line represents the isothermal curve for the constraint \eqref{eq: alpha lambda constraint} when the equality holds. Values below this line are not allowed.(b) Free energy as function of \(z_h\) and \(T\), for the cases \(T<T_c\), \(T=T_c\) and \(T>T_c\), using \(n=2\), \(L^2 V_2/k^2=1\), \(\lambda_1 L = 0\) and \(\lambda_2=1\) for simplicity.}
        \end{figure}

    Note the critical temperature \eqref{eq: critical temperature} exhibits the same shift due to the presence of \(\lambda_1 L\) as \(T_0\) in \eqref{eq: T0 temperature}. The constraint \eqref{eq: alpha lambda constraint} avoids negative values of \(T_0\).  
    The distribution of the values of \(T_c\) with respected to the parameters \(\lambda_1 L\) and \(\lambda_2\) is shown in the Fig. \ref{fig: Critical temperature distribution}.

  It is important to note that Eqs. \eqref{eq: z0 definition} and 
  \eqref{eq: critical position} imply that \(z_c<z_0\) while   
  Eqs. \eqref{eq: T0 temperature} and \eqref{eq: critical temperature} imply that $ T_0<T_c $. This result is expected from the analysis of stability .

An interesting point is that, in the limit of high dimensional spaces (i.e. large \(n\)) the pre-factors in Eqs. \eqref{eq: z0 definition} and \eqref{eq: T0 temperature}  behave, respectively, as
        \begin{align}
          & \sqrt{\frac{n+1}{n-1}}= 1 + O(1/n);\\
          & \sqrt{n^2-1}=n -  O(1/n)\, \,,
        \end{align}
In this limit \(z_0\) and \(T_0\) approach \(z_c\) and \(T_c\) respectively. 
        
The cases of more interest are the critical temperatures 
        \begin{equation}
            T_c^{(4)} = \frac{1}{\pi}\sqrt{\lambda_2}  \,+\frac{\lambda_1 L}{4\pi} \quad \mbox{and} \quad T_c^{(5)} = \frac{3}{2\pi}\sqrt{\lambda_2}  \,+\frac{\lambda_1 L}{4\pi} \,,
        \end{equation}
        corresponding to \(4D\) (\(n=2\)) and \(5D\) (\(n=3\)) spaces, respectively.

        To gain some insight into the thermodynamics of this transition, following \cite{Banerjee:2010ve}, let's construct the free energy as 
        \begin{equation}\label{eq: free energy definition}
            {\cal F} = \langle E\rangle-T S\,.
        \end{equation}
         
        Substituting the internal energy  \eqref{eq: MG internal energy} and the entropy  \eqref{eq: MG entropy} into the free energy definition \eqref{eq: free energy definition} yields:
        \begin{equation}\label{eq: MG free energy}
            {\cal F} = \frac{L^n}{2\kappa^2}V_n \left[ \frac{n}{z_h^{n+1}}+\frac{n\lambda_2}{z_h^{n-1}}+\frac{\lambda_1 L}{z_h^{n}}-\frac{4\pi}{z_h^{n}}T\right].
        \end{equation}    
        At this point we consider $z_h$ and $T $ as independent variables. The relation between them will be recovered by the extremization of ${\cal F}$. Given that the free energy in \eqref{eq: MG free energy} is expressed as a power series in \(z_h\), we can analyze it using Landau's theory of phase transitions.  Using the change of variable \(\phi = \tfrac{1}{z_h}\) for simplicity, the free energy is rewritten  as
        \begin{equation}\label{eq: MG free energy as function of phi}
            {\cal F} = \frac{L^n}{2\kappa^2}V_n \left[ n\phi^{n+1}+n\lambda_2\phi^{n-1}+\lambda_1 L \phi^n - 4\pi\phi^n T\right].
        \end{equation}
        
        Thermodynamic equilibrium requires that stable states correspond to minima of the free energy. This extremal condition yields
        \begin{equation}
            \frac{\partial}{\partial\phi}{\cal F} =n\bar\phi^n\left[\p{n+1} +\p{n-1}\lambda_2\bar\phi^{-2} + \lambda_1 L \bar\phi^{-1}-4\pi T \bar\phi^{-1}\right]=0\,,\label{derivative}
        \end{equation}
        the \(\bar \phi\) denotes the value at a minimum of \({\cal F}\). For $ n > 1 $ we find the trivial solution 
        \begin{equation}
            \bar\phi_1=0 
        \end{equation}
        that represents the space without a black hole (the \(z_h\to\infty\) limit). Using this trivial solution in \eqref{eq: MG free energy as function of phi} we find that \({\cal F}=0\) for this space.

        The non-trivial solution to \eqref{eq: MG free energy as function of phi} is derived from the following equation 
        \begin{equation}
            \p{n+1}\bar\phi_2^2  + \p{\lambda_1 L -4\pi T}\bar\phi_2 + \p{n-1}\lambda_2=0. 
        \end{equation}
        It is important to note that this equation can be directly obtained from \eqref{eq: MG temperature}, such that its solutions correspond to the inverse of the horizon position as a function of the temperature. This solution is given by
        \begin{equation}
            \bar\phi_2=\frac{-\p{\lambda_1 L - 4\pi T}\pm\sqrt{\p{\lambda_1 L-4\pi T}^2-4\p{n^2-1}\lambda_2}}{2\p{n+1}}.
        \end{equation}

        The negative sign in the root square term corresponds to \(\bar\phi_2\le\phi_0\equiv{z_0^{-1}}\), representing unstable black hole configurations. The physically relevant stable minimum, describing the black hole space,  is the one with positive sign. As it is evident from the first curve in  Fig. \ref{fig: landau like free energy}, this solution leads to \({\cal F}>0\) for temperatures \(T<T_c\), indicating that the dominant minimum corresponds to \(\bar \phi_1\), the black hole free space. For \(T>T_c\), the solution \(\bar\phi_2\) leads to \({\cal F}<0\), becoming the thermodynamically favored configuration, as shown in the last curve in Fig. \ref{fig: landau like free energy}.

        This analysis confirms our previous conclusions, obtained from analysis of the regularized actions. The phase transition occurs at \(T=T_c\), where the minima of \(\bar\phi_1\) and \(\bar\phi_2\) are equal, as seen in the purple line on Fig. \ref{fig: landau like free energy}. At this critical temperature, \(\bar\phi\) exhibits a discontinuous jump, the characteristic behavior of an order parameter in a first-order transition.

    \subsection{Case $ n=1 $ }

         In a \(3\)-dimensional space (\(n=1\)), only the polynomial \({\cal U}_1\) is non-vanishing when considering matrices in the form \eqref{eq: matrix k definition}. Therefore, in this case, we have only a single parameter \(\lambda_1\).  Thus, the action \eqref{eq: ads graviton action} reduces to 
        \begin{equation}\label{eq: 3D action }
            {\cal I} = -\frac{1}{2\kappa^2}\int d^3x\sqrt{ \vert g \vert }\p{{\cal R}+\frac{2}{L^2}}-\frac{1}{2\kappa^2}\int d^3x\sqrt{ \vert g \vert}\,\lambda_1\Tr{\cal K}\,,
        \end{equation}
   which leads to a metric of the same form as the \(AdS\) black hole, but with the horizon function now  expressed as  (see Appendix \ref{sec:Ap einstein equations})
        \begin{equation}
            f(z) = 1+\lambda_1 L z -\p{1+\lambda_1 L z}\frac{z^2}{z_h^2},
        \end{equation}
        yielding a Hawking temperature given by
        \begin{equation}\label{eq: 3D temperature}
            T=\frac{1}{4\pi z_h}\p{2+\lambda_1 L z_h}.
        \end{equation}

        Note that this temperature exhibits a similar  behavior as in the planar \(AdS\) case, but  with a \(\tfrac{\lambda_1 L}{4\pi}\) shift. That is, the temperature  decreases continuously, having a minimum value \(\lambda_1L/4\pi\) at infinity. Consequently, the black hole is stable for any finite value of \(z_h\). This can be seen by setting \(n=1\) in \eqref{eq: z0 definition}, which leads to \(z_0\to\infty\).

        Performing the same calculations as in previous sections, we obtain a regularized action expressed as 
        \begin{equation}
            {\cal I} = -\frac{V_1 L}{2\kappa^2}\beta\frac{1}{z_h^2},\label{actionn=1}
        \end{equation}
        which is the same as the \(AdS_3\) action in equation \eqref{eq: action to planar ads}.  

The free energy corresponding to action \eqref{actionn=1} is obtained from  \eqref{eq: MG free energy as function of phi} using $ n= 1$ and $\lambda_2 = 0$. In this case there is just one minimum
\begin{equation}
  \bar\phi  = \frac{4\pi T - \lambda_1 L}{2} \,        
        \end{equation}
that corresponds to the black hole solution. Since only the BH solution is stable, there is no HP transition. From this discussion, one observes that the lowest dimension in which the massive gravity system exhibits a Hawking-Page transition at non-vanishing temperature  is \(D=4\) ($n=2$).

\section{\label{sec: conclusion} Conclusion}
 
    In this work, we considered a formulation of  massive gravity like the ones studied in \cite{Hassan:2011vm,Adams:2014vza,deRham:2010ik,deRham:2010kj,Hassan:2011tf}. Since this formulation breaks the  diffeomorphism invariance \cite{Vegh:2013sk,Blake:2013bqa}, it is used in order to construct  a holographic description of systems without momentum conservation in DC conductivity study. 

    Reference \cite{Adams:2014vza} presents an investigation of  the Hawking-Page transition in an \(AdS_4\) space  with an \(S^2 \times S^1 \) boundary, within massive gravity context. It shows that the higher the graviton mass,  the lower is the critical temperature.     

 In most of the applications of holography, the manifolds involved have a planar spatial boundary. This was the main motivation for the present work, were we extended the analysis of \cite{Adams:2014vza} to the case 
 of a $\mathbb{R}^n\times\,  $S$^1 $ boundary. Our main result is given in equation \eqref{eq: critical temperature}, where we find a non-vanishing critical temperature, proportional to the parameter \(\lambda_2\), which encodes the graviton mass. This shows that in massive gravity one finds a Hawking-Page transition even in the case of a flat spatial boundary.

An interesting point to be remarked is that  the presence of a massive graviton makes the black hole unstable within a certain temperature range. This thermodynamically instability was carefully examined, and we obtained the minimum temperature \(T_0\) for the black hole to exist, highlighting the role of dimensionality in this behavior. 

We also discussed the particular three-dimensional case, which does not exhibit a Hawking-Page transition. The black hole is always stable. Consequently, we identify \(d=4\) as the lowest critical dimension for the transition occurs.

\hspace{\baselineskip}

\noindent {\bf Acknowledgments:} The authors are partially supported by CNPq --- Conselho Nacional de Desenvolvimento Científico e Tecnológico, by FAPERJ --- Fundação Carlos Chagas Filho de Amparo à Pesquisa do Estado do Rio de Janeiro, and by  Coordenação de Aperfeiçoamento de Pessoal de Nível Superior --- Brasil (CAPES), Finance Code 001.

\appendix
\section{\label{sec:Ap einstein equations} Einstein's Equations}

    Taking the variation of the action \eqref{eq: ads graviton action} with respect to the metric \(g^{mn}\) gives the following Einstein's Equations \cite{Vegh:2013sk}
    \begin{equation}\label{eq: einsteing equation to MG}
        {\cal R}_{mn} -\frac{1}{2}{\cal R}g_{mn}+\frac{1}{2}\Lambda g_{mn} + \lambda_1 {A}_{mn}+\lambda_2 L_{mn} = 0
    \end{equation}
    where \({\cal R}_{mn}\) is the Ricci tensor and 
    \begin{equation}
        A_{mn} = -\frac{1}{2}\p{g_{mn}\Tr{\cal K}-{\cal K}_{mn}} \,, \label{Amn}
    \end{equation}
    
    \begin{equation}
        L_{mn}=-\p{{\cal K}^2_{mn}-{\cal K}_{mn}\Tr{\cal K}+\frac{1}{2}g_{mn}\left[\p{\Tr{\cal K}}^2- \Tr{\cal K}^2\right] } \label{Lmn}
    \end{equation}
    are parts of the stress-energy tensor of the graviton. The tensors \eqref{Amn} and \eqref{Lmn} are obtained using the definition of \({\cal K}\) in equation \eqref{eq: matrix k definition} and writing the variations:
    \begin{align}
        &\delta\Tr{\cal K} = \frac{\partial}{\partial g^{mn}}\p{g^{mn}f_{nm}}^{1/2}\delta g^{mn}=\frac{1}{2}{\cal K}_{mn}\delta g^{mn}\,,\\
        &\delta \p{\Tr{\cal K}}^2= {\cal K}_{mn}\Tr{\cal K}\delta g^{mn}\,,\\
        &\delta \Tr{\cal K}^2 = \frac{\partial}{\partial g^{mn}} \left[\p{g^{ik}f_{kq}}^{1/2}\p{g^{qk}f_{ki}}^{1/2} \right]\delta g^{mn}={\cal K}^2_{mn}\delta g^{mn}\,.
    \end{align}

 
    Choosing metric ansatz with Poincaré \(AdS\) form
    \begin{equation}
        ds^2 =  \frac{L^2}{z^2}\p{-f(z)dt^2+f(z)^{-1}dz^2+d\vec x^2} \,,
    \end{equation}
     and using \eqref{eq: refence metric}. Eq. \eqref{eq: einsteing equation to MG} leads to the following linear differential equation for the \(t t\) component: 
    \begin{equation}
        z f'(z)-\p{n+1}f(z)+\p{n+1}+\lambda_1L z+\lambda_2 \p{n-1}z^2=0\,.
    \end{equation}
    
    Solving for \(f(z)\) one finds 
    \begin{equation}\label{eq: integral equation to f}
        f(z) = z^{n+1}\left[\int dz \p{-\frac{\p{n+1}}{z^{n+2}}-\frac{\lambda_1 L}{z^{n+1}}-\frac{\lambda_2\p{n-1}}{z^n}}+C\right].
    \end{equation}
    
    At this point, note that for \(n=1\), the \(\lambda_2\)-term vanishes. This occurs because, for a \(3\times 3\) matrix as defined in \eqref{eq: matrix k definition}, only the \({\cal U}_1=\Tr{\cal K}\) symmetric polynomial is non-zero. 
    
    First, let us consider the solution for the case \(n>1\), which reads 
    \begin{equation}\label{eq: generic f}
        f(z) = 1 + \frac{\lambda_1 L}{n}z+\lambda_2 z^2  - M z^{n+1} \,.
    \end{equation}
    
    The integration constant \(C\) has been redefined as \(-M\), which can be determined requiring that \(f(z_h)=0\). Thus,
    \begin{equation}
        M = \frac{1}{z^{n+1}}\p{1+\frac{\lambda_1 L z_h}{n}+\lambda_2 z_h^2}
    \end{equation}
    indicating that \(M\) is proportional to the black hole's energy density (see \eqref{eq: MG internal energy}). Substituting this result back into \eqref{eq: generic f}, one finds the horizon function \eqref{eq: MG horizon function}.

    For the case \(n=1\), the horizon function follows from equation \eqref{eq: integral equation to f} is 
    \begin{equation}
        f(z) = 1+ \lambda_1 L z-\p{1+\lambda_1 L z_h}\frac{z^2}{z^2_h}
    \end{equation}
    where we already imposed the condition \(f(z_h)=0\). Setting \(\lambda_1=0\), we recover the usual \(AdS_3\) black hole solution, as can be seen in \cite{dosSantos:2022scy}.
    

\bibliography{mybib}

\end{document}